\documentclass[journal,11pt]{IEEEtran}
\usepackage{latexsym}
\usepackage{graphicx}
\usepackage{amsfonts,amssymb,amsmath}
\usepackage{mathtools, cuted}
\usepackage{hyperref}
\usepackage{amsmath}
\usepackage[T1]{fontenc}
\usepackage{cite}
\usepackage{subcaption}
\usepackage{comment}
\usepackage{amsthm}
\usepackage{multirow}
\usepackage{overpic}
\usepackage{steinmetz}
\usepackage{array}
\usepackage{url}
\usepackage{color}
\usepackage{tikz}

\usepackage{algorithm}
\usepackage{algorithmic}
\usepackage{units}
\usepackage{mathrsfs}

\theoremstyle{plain}

\newcommand{\vect}[1]{\mathbf{#1}}

\def\tr{\mathrm{tr}}

\def\Htran{\mbox{\tiny $\mathrm{H}$}}
\def\Ttran{\mbox{\tiny $\mathrm{T}$}}
\def\imagunit{\mathsf{j}} 
\def\SNR{\mbox{\footnotesize $\mathrm{SNR}$}}

\title{Exploiting the Depth and Angular Domains for Massive Near-Field Spatial Multiplexing}

\author{Parisa Ramezani, Alva Kosasih, Amna Irshad, and Emil Bj{\"o}rnson}

\begin{document}

\maketitle

\begin{abstract}
In this article, we present our vision for how extremely large aperture arrays (ELAAs), equipped with hundreds or thousands of antennas, can play a major role in future 6G networks by enabling a remarkable increase in data rates through spatial multiplexing of a massive number of data streams to both a single user and many simultaneous users. Specifically, with the quantum leap in the array aperture size, the users will be in the so-called radiative near-field region of the array, where previously negligible physical phenomena dominate the propagation conditions and give the channel matrices more favorable properties. This article presents the foundational properties of communication in the radiative near-field region and then exemplifies how these properties enable two unprecedented spatial multiplexing schemes: depth-domain multiplexing of multiple users and angular multiplexing of data streams to a single user. We also highlight research challenges and open problems that require further investigation. 
\end{abstract}

\vspace{-4mm}

\section{Introduction}

The data traffic is growing at an exponential pace in wireless communication systems. To cater to this development, we must constantly increase the total bit rate (in bits per second) that the systems provide. The upper rate limit is the Shannon capacity
\begin{equation} \label{eq_basic_capacity}
B \log_2 \Big( 1 + \underbrace{\frac{P \beta}{B N_0}}_{= \SNR} \Big),
\end{equation}
which is determined by the transmit power $P$, channel gain $\beta \in [0,1]$, and noise power spectral density $N_0$.
From inspecting \eqref{eq_basic_capacity}, it appears that increasing the bandwidth $B$ is the preferred way to enhance capacity.
The signal-to-noise ratio (SNR) can also be improved through beamforming, but the impact is smaller due to the logarithm.
Hence, the bandwidth has grown with every cellular network generation, with 100 MHz being typical in the first phase of 5G deployments. 
Since spectrum is a scarce global resource, adding new bands is generally associated with moving to higher frequencies. For example, the second phase of 5G will use mmWave bands in the range 24-53 GHz \cite{Parkvall2017a}, while 6G is expected to also make use of the sub-THz band spanning from 100 to 300 GHz \cite{Rappaport2019a}.

\begin{figure}[t!]
       \centering
       \begin{overpic}[width=\columnwidth,tics=10]{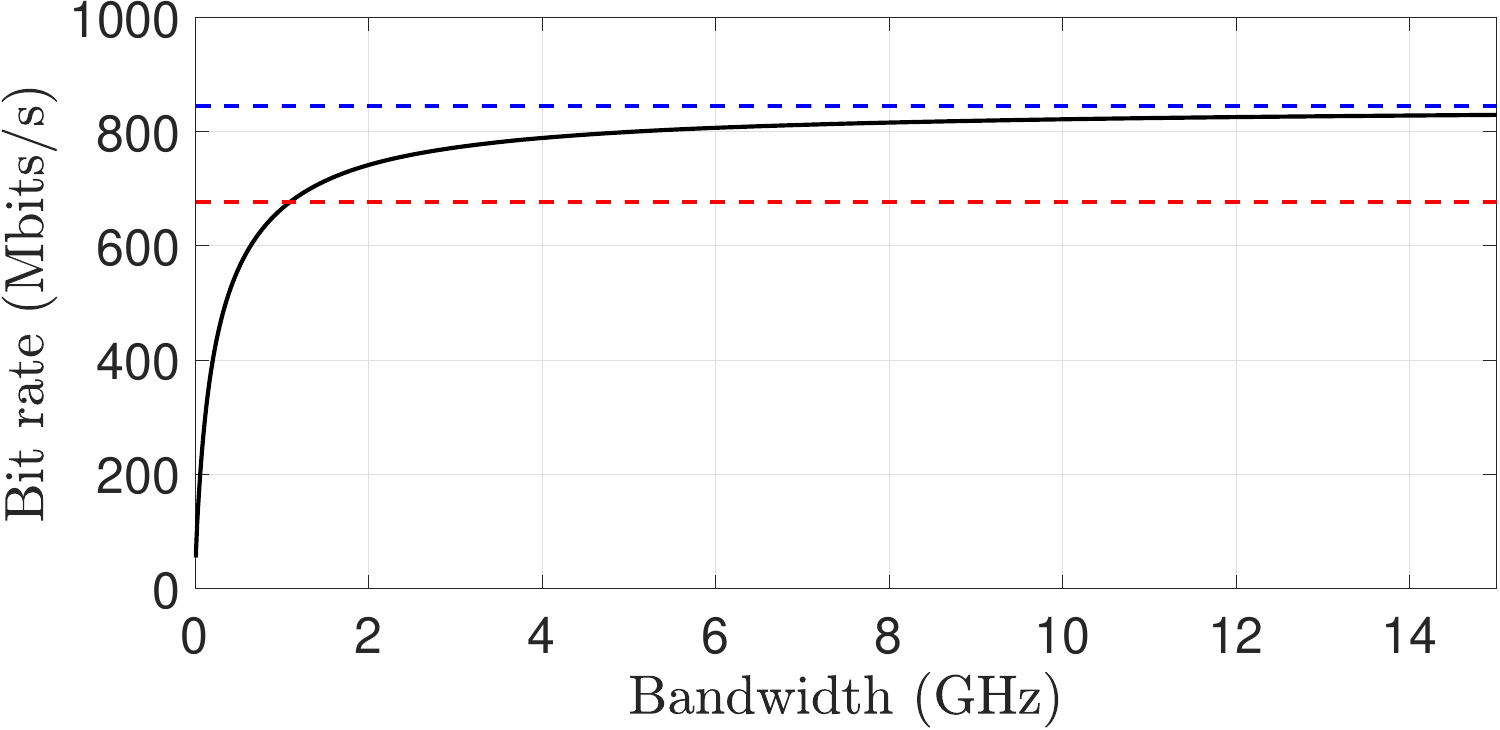}
  \put(55,30.4){\small 80\% of upper limit}
  \put(67.5,43){\small Upper limit}
  \put(23,23){\small Linear growth}
\put(23,27){\vector(-1,1){5}}
\end{overpic}  \vspace{-4mm}
       \caption{The bit rate of communication link grows with the bandwidth but approaches an upper limit. The received power $P \beta $ is fixed for a 400 m line-of-sight channel. The upper limit $\log_2(e) P\beta /N_0$ is approached as $B \to \infty$.}
       \label{fig:convergence_upper_bound}
   \end{figure}
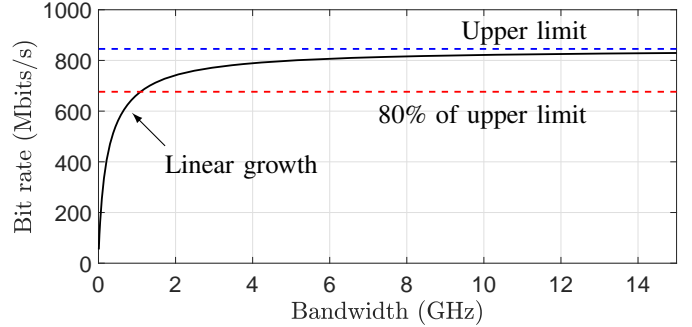

These new bands feature gradually worse channel gain conditions, and hardware limitations prevent the transmit power from increasing proportionally to the increased bandwidth.
We can partially compensate for this by using more antennas for extremely narrow beamforming \cite{Rappaport2019a}.
However, even if we design advanced beamforming systems to reach the same received power levels in these new bands as in the current ones, the extra bandwidth gives diminishing returns since we must divide the signal power over it. 
Fig.~\ref{fig:convergence_upper_bound} shows how the bit rate grows almost linearly with the bandwidth up to $1$ GHz and then approaches an upper bound. We will reach this saturation level in the 5G era. At very short distances, we might reach saturation in 6G using around $10$ GHz of spectrum in sub-THz bands. The main point is that the spectrum resource will eventually be depleted, and we need to look for new design dimensions to keep raising the capacity.

\subsection{A Vision for Massive Spatial Multiplexing}

How can we continue increasing the bit rate when the bandwidth resource is depleted? The only known alternative way is to transmit multiple parallel data streams as \emph{spatial layers} \cite{Anderson1991a,Telatar1999a,Foschini1998a}.
This approach requires antenna arrays at both the transmitter and receiver sides, and is known as \emph{multiple-input multiple-output (MIMO)}. It comes in two configurations.
In single-user MIMO (SU-MIMO), both the base station (BS) and user equipment (UE) are equipped with antenna arrays so that parallel spatial layers can be transmitted between them, using different angular dimensions. In multi-user MIMO (MU-MIMO), the BS communicates simultaneously with multiple UEs, each having one or a few antennas.
Both features are used in 5G \cite{Parkvall2017a}, but the number of spatial layers is limited to $8$ (per polarization). Unfortunately, it is hard to reach even that many layers in practice. The point-to-point channels in SU-MIMO mode seldom provide more than $1$-$2$ strong propagation paths that can carry different data layers. Moreover, the UEs typically have so similar channels in MU-MIMO mode that one must over-provision the BS with antennas to limit the interference (e.g., using $64$ antennas to serve $8$ UEs). This is called the massive MIMO regime \cite{Marzetta2010a,massivemimobook} and 5G is built around it.
Does this mean that the spatial layering resource is also about to be depleted?

No, we envision that we still have the \emph{massive spatial multiplexing} era ahead of us. By using physically larger arrays (at the BSs) and smaller wavelengths, future communication systems can operate in the \emph{radiative near-field} where previously negligible physical phenomena become dominant. 
In this article, we will show how these phenomena enable enhanced spatial multiplexing where each spatial layer has an unprecedentedly small focus area. The motivation goes back to the fundamental MIMO capacity expression \cite{Telatar1999a}
\begin{equation} \label{eq:MIMO_capacity}
C =  \max_{\vect{Q}: \, \tr(\vect{Q}) \leq 1 } B \log_2 \det ( \vect{I}_K + \SNR \cdot \vect{H} \vect{Q} \vect{H}^{\Htran}),
\end{equation}
where $\vect{H} \in \mathbb{C}^{K \times K}$ is the channel matrix between $K$ transmit antennas and $K$ receive antennas, and $\vect{Q}$ is the covariance matrix of the transmitted signal. For a given channel matrix, it is well known that the maximum in \eqref{eq:MIMO_capacity} is achieved by transmitting along the right singular vectors of $\vect{H}$ and dividing the power using waterfilling \cite{Telatar1999a}.
However, the practical challenge is to design future systems so that the channel matrix has favorable properties. The MIMO capacity in \eqref{eq:MIMO_capacity} is a function of $\vect{H}$ and is maximized, under the Frobenius norm constraint $\| \vect{H} \|^2_\mathrm{F}=K^2$, when all its $K$ singular values are equal to $K$. In this case, $\vect{Q} = \frac{1}{K} \vect{I}_K$ is the optimal signal covariance matrix. 
It follows that \eqref{eq:MIMO_capacity} is upper bounded as
\begin{equation} \label{eq:upper_limit}
C \leq B K \log_2 ( 1 + \SNR),
\end{equation}
where the bound is an increasing function of $K$, which we refer to as the \emph{multiplexing gain}. Hence, if we can achieve channel matrices of the aforementioned kind, we can manage the anticipated traffic growth by increasing the number of spatial layers (and the number of transmit/receive antennas) instead of using more spectrum.

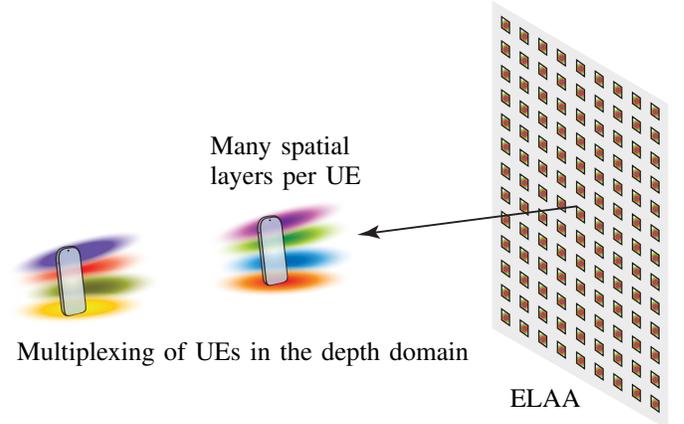
\begin{figure}[t!]
       \centering
       \begin{overpic}[width=\columnwidth,tics=10]{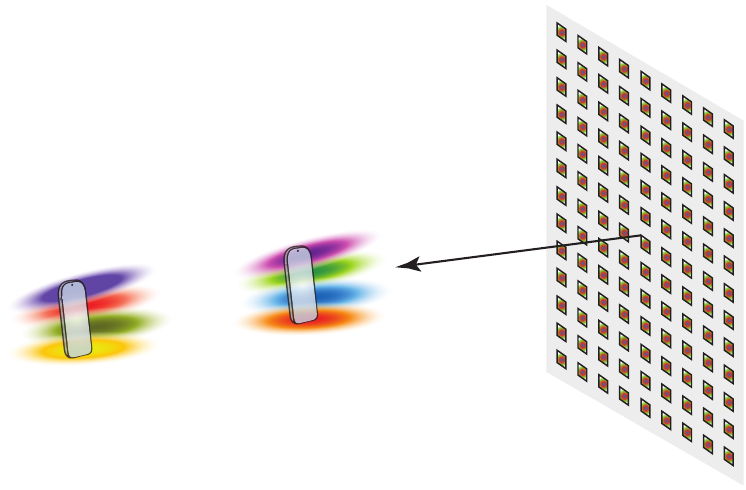}
  \put(76,4){\small ELAA}
  \put(31,42){\small Many spatial}
  \put(31,37.5){\small layers per UE}
  \put(2,11){\small Multiplexing of UEs in the depth domain}
\end{overpic}  \vspace{-6mm}
       \caption{A BS equipped with an ELAA has an unprecedented spatial resolution that enables spatial multiplexing of UEs in the depth domain and sending multiple layers per UE, with low interference even in LOS scenarios.}
       \label{fig:multi_layer}
       \vspace{-3mm}
   \end{figure}

In the remainder of this article, we will first outline the previously overlooked near-field propagation phenomena and then showcase how these enable us to approach \eqref{eq:upper_limit} when $K$ is large, in both SU-MIMO and MU-MIMO scenarios. For brevity in the presentation, we will focus on line-of-sight (LOS) channels, for which the results are easiest to interpret.
We will refer to the envisioned BS technology as an \emph{extremely large aperture array (ELAA)} \cite{Bjornson1}, which is a term highlighting that the array is large compared to the wavelength. An ELAA can be physically large or small, depending on the wavelength of operation. 
Fig.~\ref{fig:multi_layer} gives a first schematic description of how an ELAA can both transmit multiple spatial layers per UE and multiplex UEs in the depth domain. The technical details will follow.

\section{Near-Field Properties}
A transmit antenna consists of point sources that emit spherical waves. A point source at the location $\vect{s} \in \mathbb{R}^3$ generates an electric field at an observation point located at $\vect{r} \in \mathbb{R}^3$, given by the tensor Green’s function as \cite{Poon}
\begin{equation}
\begin{aligned}
\label{eq:green_function}
    G(d) &= - \frac{j\eta e^{-j \frac{2\pi}{\lambda} d}}{2\lambda d} \Big[ (\vect{I} - \hat{\vect{d}}\hat{\vect{d}}^{\Htran}) + \frac{j\lambda}{2\pi d} (\vect{I} -3 \hat{\vect{d}}\hat{\vect{d}}^{\Htran})\\ 
    &- \frac{\lambda^2}{(2\pi d)^2}(\vect{I} -3 \hat{\vect{d}}\hat{\vect{d}}^{\Htran})\Big],
    \end{aligned}
\end{equation}
where $d = \| \vect{r} - \vect{s}\|$ is the distance between the source and the observation point, $\eta$ is the free space impedance, $j$ is the imaginary number, $\lambda$ is the wavelength, and $\hat{\vect{d}}= (\vect{r} - \vect{s})/d$ is a unit-length vector denoting the direction of $d$. The power of the last two terms in \eqref{eq:green_function} decays rapidly with $d$; thus, they are only considered when characterizing the electric field in the \emph{reactive near-field region}, very close to the antenna \cite{Bjornson4}. The reactive near-field starts from the surface of the antenna and continues to $d = d_{\mathrm{N}}$. For electrically small antennas, it is common to assume $d_{\mathrm{N}} = \lambda/(2\pi)$, although $d_{\mathrm{N}} = \lambda$ has been experimentally verified to be a better choice \cite{Yaghjian}. The reactive near-field ends approximately at $d_{\mathrm{N}} = 0.62\sqrt{D^3/\lambda}$ for electrically large antennas, where $D$ is the antenna's largest dimension \cite{Selvan}.

The radiative near-field begins after the reactive near-field and is traditionally said to cover distances between $d_\mathrm{N}$ and $d_{\mathrm{F}}$ from the antenna (i.e., $d_{\mathrm{N}} < d <d_{\mathrm{F}}$). The distance at which the radiative near-field ends and the far-field begins is called the Fraunhofer distance \cite{Cheng,Sherman} or the Rayleigh distance \cite{Kraus2002} and is commonly taken as 
\begin{equation}
\label{eq:Fraun_dist}
    d_{\mathrm{F}} = \frac{2D^2}{\lambda}.
\end{equation}
This is the distance beyond which the radiated spherical waves can be viewed as approximately planar, in the sense that the phase difference between the waves from the antenna's center and edge is smaller than
$\pi/8$ at the observation point.
The Fraunhofer distance is typically very short, even for relatively large antenna apertures. As an example, for an antenna with the maximum length of $D=2\lambda$ operating at $f=3$ GHz, the Fraunhofer distance is computed as $d_{\mathrm{F}} = 8\lambda = \frac{8c}{f} = 0.8$ m, where $c = 3 \cdot 10^8$ m/s is the speed of light.
Hence, the radiative near-field region has traditionally been overlooked by the communication community, where the ``near-field'' term has been reserved for short-range systems that utilize inductive coupling in the reactive near-field. This perception will likely change in the 6G era.

\subsection{Fraunhofer Array Distance}

\begin{figure}[t!]
       \centering \vspace{5mm}
       \begin{overpic}[width=\columnwidth,tics=10]{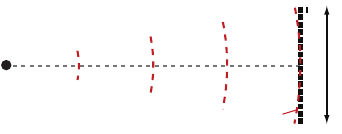}
  \put(0,29){\small Transmitting}
  \put(0,24.5){\small point source}
  \put(57,4){\small Spherical wave}
  \put(67,39){\small Receiving ELAA}
  \put(50,21){\small $d$}
  \put(90.5,34.5){\small $D$}
  \put(96.5,19){\small $W$}
\end{overpic}  \vspace{-6mm}
       \caption{When a user device (point source) communicates with an ELAA, the propagation distance $d$ typically satisfies $d_{\mathrm{F}}<d<d_{\mathrm{FA}}$. This implies that each antenna is in the far-field, but the array is in the radiative near-field, so it can resolve spherical wavefronts and utilize them.}
       \label{fig:figure_fraunhofer_array}
   \end{figure}
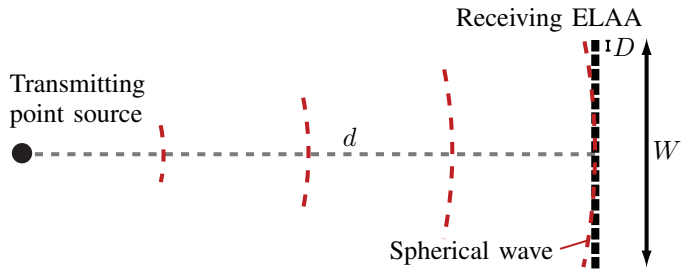

If current antenna arrays evolve into ELAAs, the radiative near-field gets a new meaning. A large array contains many antennas that each has a small aperture $D$, but the maximum dimension $W$ of the array is much larger. We will consider  planar arrays where $W$ is its diagonal, but the specific details are given later. This creates the new situation illustrated in Fig.~\ref{fig:figure_fraunhofer_array}. We have exchanged the roles of the transmitter and receiver for a simplified presentation but stress that the channel reciprocity implies that the channel properties are the same in both directions.
The point source emits a spherical wave, and when it reaches the antenna array, each antenna will observe a locally plane wave because the propagation distance $d$ satisfies $d > d_{\mathrm{F}}$. The spherical curvature is, however, noticeable when comparing the received signals at different antennas in the array. This property must be taken into account when modeling the channel vector.

The term \emph{Fraunhofer array distance} has recently sprung up to quantify at what distances this phenomenon appears \cite{Bjornson4,Ramezani2022}. If we want the phase variations to be negligible (i.e., smaller than $\pi/8$) across an array with the maximum dimension of $W$, the Fraunhofer array distance becomes
\begin{equation}
\label{eq:Fraun_array_dist}
    d_{\mathrm{FA}} = \frac{2W^2}{\lambda}.
\end{equation}
This is nothing but the Fraunhofer distance in \eqref{eq:Fraun_dist} evaluated using the array aperture $W$ instead of the antenna aperture $D$. However, the consequences are very different because an array can compensate for phase variations and exploit them for new types of beamforming and multiplexing, as we will shortly discuss.

The Fraunhofer array distance for an ELAA operating at $f=3\,$GHz with the size $1.4 \times 1.4$\,m and aperture length $W = 2\,$m is $d_{\mathrm{FA}} = 80\,$m. If the same ELAA operates at $f = 30\,$GHz, the Fraunhofer array distance increases to $d_{\mathrm{FA}} = 800\,$m. If new deployment practices enable even larger arrays, then the $d_{\mathrm{FA}}$ can be several kilometers.
Hence, if we take the Fraunhofer array distance as the border between radiative near-field and far-field regions when communicating using an ELAA, many prospective users served by the BS would fall in its near-field region where the spherical wavefront of the emitted waves must be properly modeled.

\subsection{What Array Gain can be Achieved?}

Suppose the ELAA in Fig.~\ref{fig:figure_fraunhofer_array} has $N$ antennas. The received signal $\vect{y} \in \mathbb{C}^N$ can then be expressed as
\begin{equation}
    \vect{y} = \vect{h} x + \vect{n},
\end{equation}
where $\vect{h} \in \mathbb{C}^N$ is the channel vector, $x$ is the transmitted signal, and $\vect{n} \in \mathbb{C}^N$ is the receiver noise.
This looks like a conventional single-input multiple-output (SIMO) channel, but the channel vector is modeled differently. 
In a conventional far-field LOS scenario, $\vect{h}= \sqrt{\beta} [1, \ldots,1]^{\Ttran}$, where $\beta$ is the channel gain.
In the radiative near-field, there can be both phase and amplitude variations in $\vect{h}$; an exact model will be provided later.
In any case, the amplitude/phase variations can be compensated for using matched filtering:
\begin{equation}
    \frac{\vect{h}^{\Htran}}{\| \vect{h}\|}\vect{y} = \| \vect{h}\| x + \frac{\vect{h}^{\Htran}}{\| \vect{h}\|}\vect{n}.
\end{equation}
This leads to an SNR proportional to $\| \vect{h}\|^2$, which coherently combines the total received signal power over all the receive antennas.
This capability is not possessed by a single large antenna with aperture $W$, which will basically perform the filtering $[1, \ldots,1]^{\Ttran} \vect{h} / \sqrt{N}$ in the analog domain, so that the phase-shifts caused by the spherical wavefront makes the received power average out between the antennas. In other words, spherical waves are detrimental to large antennas but not to antenna arrays.

We now consider that a single-antenna isotropic  transmitter located at the location $(0,0,z)$ sends a signal to an ELAA deployed in the $xy$ plane centered at the origin.
In contrast to the one-dimensional ELAA illustrated in Fig.~\ref{fig:figure_fraunhofer_array}, we now consider a planar array that has $N$ antennas in each row and $M$ antennas in each column. 
The channel vector $\vect{h}$ is now $MN$-dimensional. If the channel gain is $\beta$ to an arbitrary antenna in the center of the ELAA, we would expect that $\| \vect{h} \|^2 = \beta MN$ when communicating in the far-field. The factor $MN$ is called the \emph{array gain}. The situation is different in the radiative near-field, where we instead get
\begin{equation}
\| \vect{h} \|^2 = \beta MN G_{\textrm{array}},
\end{equation}
where the \emph{normalized array gain} is computed as \cite{Bjornson4}
\begin{equation} \label{eq:antenna-array-gain-exact}
G_{\textrm{array}} = \frac{ \sum_{m=1}^{M} \sum_{n=1}^{N}  \left|  \int_{\mathcal{S}_{m,n}}  E(x,y) dx \, dy \right|^2 }{ MN A  \int_{\mathcal{S}} \left| E(x,y) \right|^2 dx \, dy},
\end{equation}
where $A$ is the physical area of each antenna, $\mathcal{S}_{m,n}$ is the set of points in the $xy$-plane spanned by the antenna in the $m$th row and $n$th column of the ELAA, $\mathcal{S}$ is the area spanned by the reference antenna located in the origin, and
\begin{equation} \label{eq:intensity-function}
E(x,y) = \frac{E_0}{\sqrt{4 \pi}} \frac{\sqrt{z (x^2 + z^2)}}{(x^2+y^2+z^2)^{5/4}} e^{-\imagunit \frac{2\pi}{\lambda}\sqrt{x^2+y^2+z^2}},
\end{equation} 
is the electric field measured an arbitrary point $(x,y,0)$ with $E_0$ being the electric density. The expression in \eqref{eq:antenna-array-gain-exact} is the total power received by the $MN$ antennas divided by the total received power of $MN$ reference antennas of the kind in the origin.
The same method can be used to compute the exact complex-valued channel coefficient $h_{m,n}$ in $\vect{h}$ for the antenna in the $m$th row and $n$th column:
\begin{equation}
h_{m,n} = \frac{1}{E_0} \sqrt{\frac{1}{A}} \int_{S_{m,n}} E(x,y) dx dy.
\end{equation}
The expression in \eqref{eq:antenna-array-gain-exact} becomes $G_{\mathrm{array}} = 1$ in the far-field (i.e., $z \gg 0$), where the electric field in \eqref{eq:intensity-function} can be approximated as $E_0/(\sqrt{4\pi} z)$.
However, it can be smaller in the radiative near-field, even if the receiving array is capable of compensating for the phase-shifts. When that happens, we cannot make efficient use of the entire array area.
As indicated above, the Fraunhofer array distance $d_{\mathrm{FA}} $ specifies the distance beyond which a spherical wave is seen as planar by the ELAA. Does $d_{\mathrm{FA}}$ also determine the distance where the maximum normalized array gain is achievable?

No, it has been recently reported in \cite{Bjornson4,Ramezani2022} that the array gain in \eqref{eq:antenna-array-gain-exact} is very close to $1$ in the majority of the radiative near-field. To shed light on this behavior, we will examine the planar ELAA in detail.  The maximum dimension (i.e., aperture length) of the ELAA is obtained as 
\begin{equation}
   W = D\sqrt{\frac{M^2+N^2}{2}}, 
\end{equation}where $D$ denotes the diagonal of a single antenna. The Fraunhofer array distance for the ELAA is thus given by 
\begin{equation}
\label{eq:Fraun_array_dist_ELAA}
    d_{\mathrm{FA}} = \left( \frac{W}{D}\right)^2 d_{\mathrm{F}} = \frac{M^2+N^2}{2}\, d_{\mathrm{F}}.
\end{equation} 

\begin{figure}[t!]
	\centering
	\begin{overpic}[width=\columnwidth]{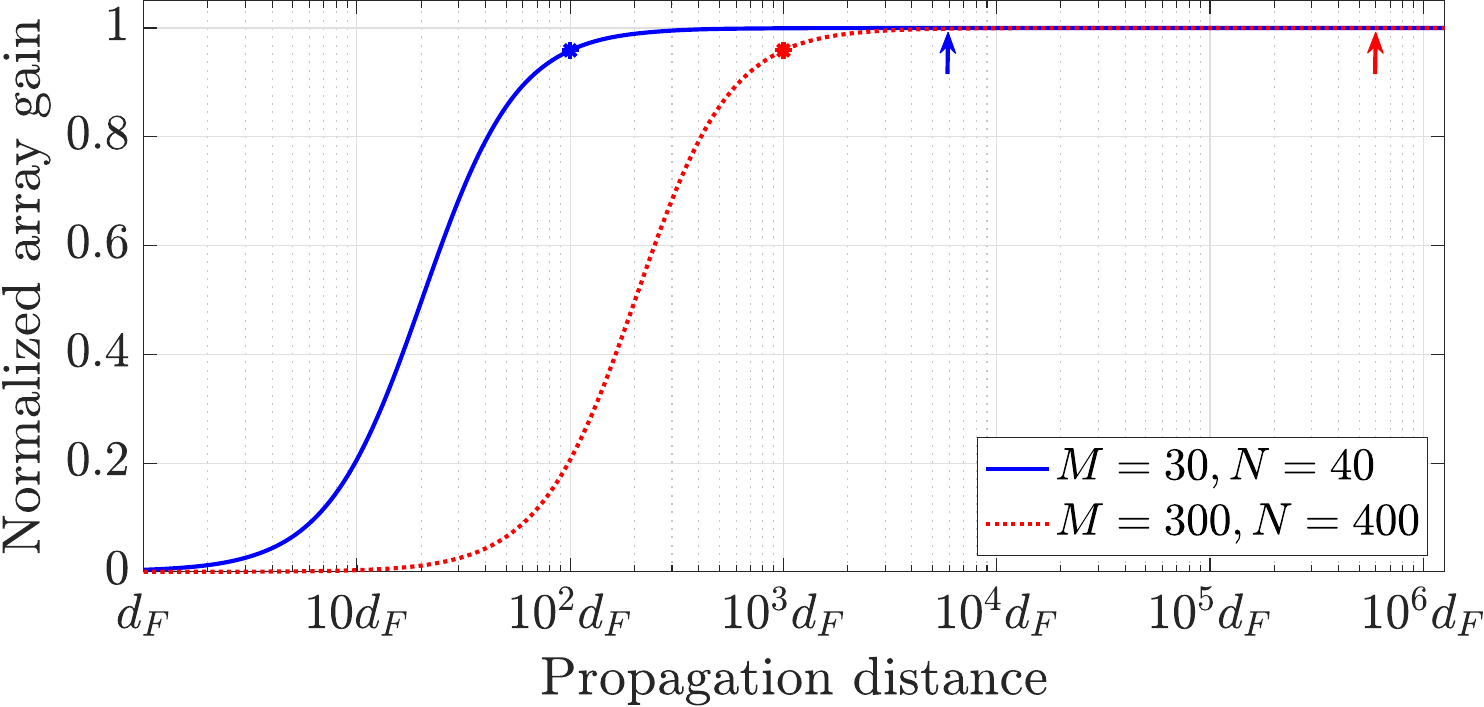}
	 \put (60,40) {$d_{\mathrm{FA}}$}
	 \put (90,40) {$d_{\mathrm{FA}}$}
\end{overpic}  
	\caption{The normalized array gain in \eqref{eq:antenna-array-gain-exact} depends on the distance in the radiative near-field, but converges to its maximum value at a distance much shorter than the Fraunhofer array distance. Hence, we can usually achieve the maximum array gain in the near-field.}
	\label{fig:nomralized_array_gain} 
\end{figure}

Fig.~\ref{fig:nomralized_array_gain} depicts the normalized array gain in \eqref{eq:antenna-array-gain-exact} for two ELAAs with different numbers of antennas. In particular, the solid blue curve shows the array gain for an ELAA with $M = 30$ rows and $N=40$ columns of antennas, while the dotted red curve represents the array gain of another ELAA with $M= 300$ and $N = 400$. Each antenna has the size $\lambda/4 \times \lambda/4$. From \eqref{eq:Fraun_array_dist_ELAA}, their Fraunhofer array distances are respectively computed as $ 1.25 \cdot 10^3 d_{\mathrm{F}}$ and $ 1.25 \cdot 10^5 d_{\mathrm{F}}$, indicated by arrows on the graphs. We note that a logarithmic scale is used on the horizontal axis. It can be observed that the normalized antenna array gains converge to $1$ much earlier than $d_{\mathrm{FA}}$, which suggests that the Fraunhofer array distance is not a good indicator of when the maximum array gain is achievable. This is an encouraging result because it demonstrates that we can achieve the maximum array gain in the vast majority of the radiative near-field, if we just utilize a matched filter receiver that compensates for the phase variations caused by the spherical waves.

For a planar square ELAA, it was first shown in \cite{Bjornson2} that almost $96\%$ of the maximum array gain is achieved for $d \geq d_{\mathrm{B}}$, where $d_{\mathrm{B}}$ is twice the largest dimension of the ELAA and characterizes when the propagation distances from the transmitter to different parts of the receiver are so large that they affect the array gain. For the setup considered above, with an unequal number of rows and columns,  this distance becomes

\begin{equation}
    d_{\mathrm{B}} = 2W = 2D\sqrt{\frac{M^2+N^2}{2}}.
\end{equation}
For the ELAAs investigated in Fig.~\ref{fig:nomralized_array_gain}, 
$d_{\mathrm{B}}$ is respectively obtained as $10^2 \,d_{\mathrm{F}}$ and $10^3 \,d_\mathrm{F}$, shown by star markers on the curves. For both graphs, we have $G_{\mathrm{array}} \approx 0.96$ at $d = d_{\mathrm{B}}$, just as in the case with square ELAAs.
\vspace{-3mm}
\subsection{Beam Width and Beam Depth}

\begin{figure}[t!]
       \centering
       \begin{overpic}[width=\columnwidth,tics=10]{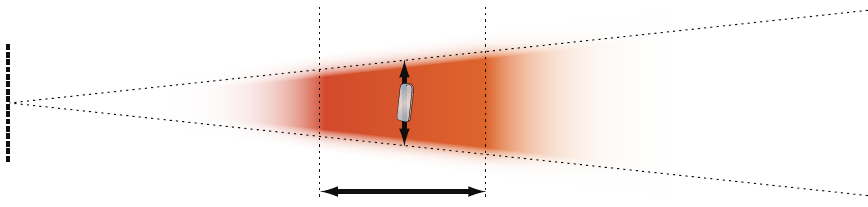}
  \put(0,1){\small ELAA}
  \put(38.5,11){\small BW}
  \put(44,3){BD}
\end{overpic}  \vspace{-4mm}
       \caption{When transmitting to a user in the radiative near-field, the array gain will be strong in a focus area around the user with limited width and depth.}
       \label{fig:beamdepth_nearfield}
   \end{figure}

When an ELAA transmitter uses matched filtering to focus the transmitted signal at a point $(0,0,F)$, the maximum array gain is obtained at that point and smaller numbers at other points $(x_r,y_r,z_r)$. 
The novel aspect is that the focus area has both a narrow beam width (BW) and beam depth (BD), where the latter is a new concept for the near-field. These properties are sketched in Fig.~\ref{fig:beamdepth_nearfield}.

The $3\,$dB BW and BD can be derived by utilizing the 
so-called Fresnel approximation of the electric field in \eqref{eq:intensity-function},
\begin{equation}
    E(x,y) \approx \frac{E_0}{\sqrt{4\pi}z}e^{-j\frac{2\pi}{\lambda}(z+\frac{x^2}{2z}+\frac{y^2}{2z})},
\end{equation}
which is tight in the part of the radiative near-field where the array gain is maximum (i.e., $d_{\mathrm{B}}<z<d_{\mathrm{FA}}$).
If the elements in the ELAA are small, matched filtering is equivalent to multiplying $E(x,y)$ with $e^{+j\frac{2\pi}{\lambda}(\frac{x^2}{2F}+\frac{y^2}{2F})}$, which is the phase-shift at the focal point. 
To compute the $3\,$dB BW, we consider a receiver located at $(x_r,y_r,F)$ for which the normalized array gain can be shown to become \cite{Bjornson4}
\begin{equation}
\label{eq:array_gain_BW}
   G_{\mathrm{array}}(x_r,y_r,F) \approx \mathrm{sinc}^2\left(\frac{N}{\sqrt{2}}\frac{Dx_r}{\lambda F}\right) \mathrm{sinc}^2\left(\frac{M}{\sqrt{2}}\frac{Dy_r}{\lambda F}\right). 
\end{equation}
Since $\mathrm{sinc}^2(x)$ has its maximum value at $x = 0$ and we have $\mathrm{sinc}^2(\pm 0.443) \approx 0.5$, the array gain in \eqref{eq:array_gain_BW} is half of its maximum value at 
\begin{equation}
\label{eq:x_3dB}
   x_{r,3\,\mathrm{dB}} \approx \pm  \frac{0.443\sqrt{2}\lambda F}{ND} ,
\end{equation} and the $3\,$dB BW along the $x$-axis can be computed as 
\begin{equation}
\label{eq:BW_3dB}
    \mathrm{BW}_{\mathrm{3\,dB}} \approx 0.886\sqrt{2}\, \frac{\lambda F}{ND}.
\end{equation}
This is the same expression as in far-field communications; in fact, the angular 3dB BW is approximately $\mathrm{BW}_{\mathrm{3\,dB}}/F$ radians and is independent of the distance to the focal point. However, the BW in meters becomes much smaller when the array is large; thus, the beams are narrower in the radiative near-field.

\begin{figure}[t!]
       \centering
       \includegraphics[width=\columnwidth]{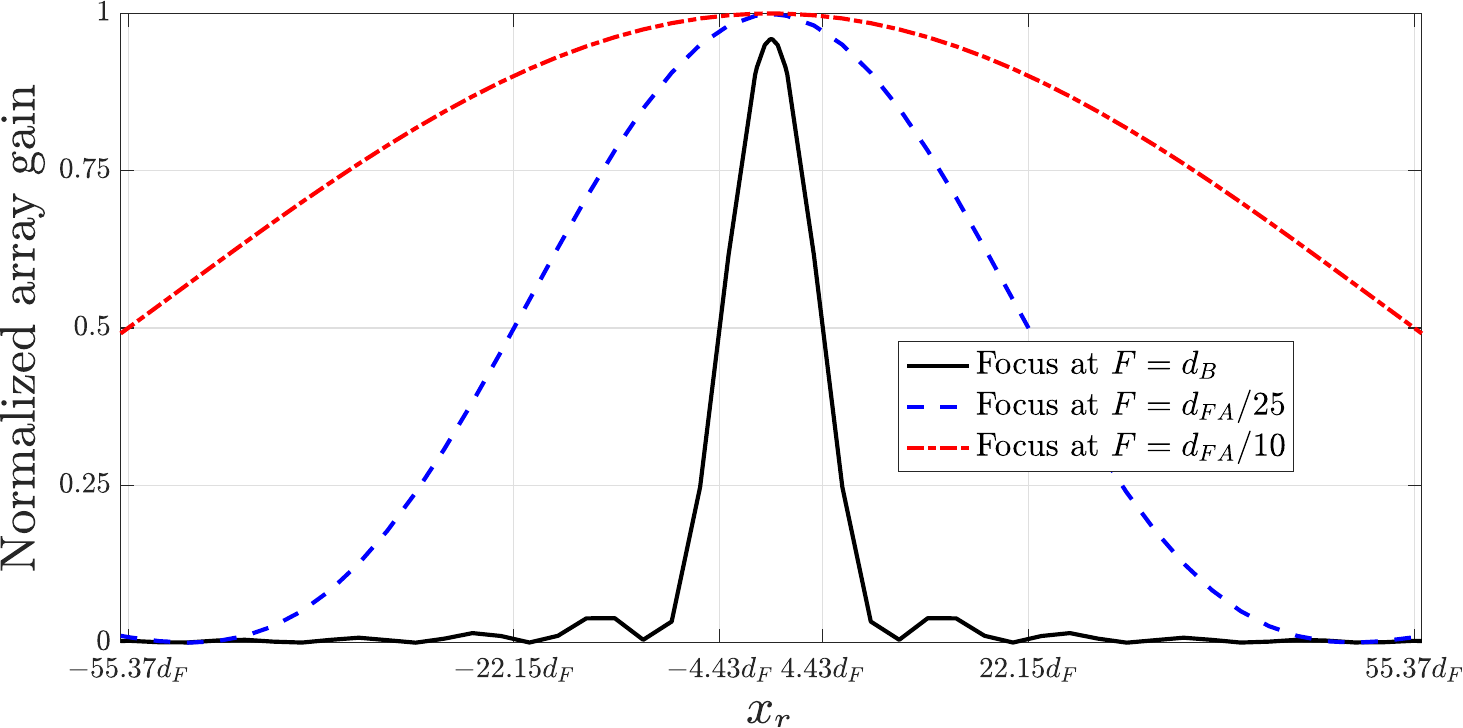}
       \caption{Normalized array gain vs. $x$ coordinate of the receiver for an ELAA transmitter with $N=400$ elements in each row. The $3$\,dB BW increases when the ELAA focuses at further distances.}
       \label{fig:beam_width}
   \end{figure}

Fig.~\ref{fig:beam_width} illustrates the normalized array gain of an ELAA with $M=300$ and $N=400$, consisting of antennas of the size $\lambda/4 \times \lambda/4$. The array gain is shown as a function of the $x$ coordinate of a receiver located at $(x_r,0,F)$. We consider three focusing scenarios: $F = d_{\mathrm{B}} = 10^3\,d_{\mathrm{F}}$, $F = d_{\mathrm{FA}}/25 = 5\cdot 10^3 \, d_{\mathrm{F}}$, and $F = d_{\mathrm{FA}}/10 =1.25 \cdot 10^4\, d_\mathrm{F} $. It can be observed that the normalized array gain is approximately $0.5$ when $x_r = \pm 4.43\, d_{\mathrm{F}}$ for $F = d_{\mathrm{B}}$ which can also be obtained from \eqref{eq:x_3dB}. 
The $3\,$dB BW can accordingly be obtained as $\mathrm{BW}_{\mathrm{3\,dB}} = 8.86\, d_{\mathrm{F}}$. This number increases to $\mathrm{BW}_{\mathrm{3\,dB}} = 44.3\, d_{\mathrm{F}}$ and $\mathrm{BW}_{\mathrm{3\,dB}} = 110.75 \, d_{\mathrm{F}}$ for $F = d_{\mathrm{FA}}/25$ and $F = d_{\mathrm{FA}}/10$, respectively.

To characterize the BD, we instead compute the normalized array gain for a receiver placed at $(0,0,z_r)$, in the same direction as the focal point for the beam but at a different distance. The normalized array gain can then be shown to become \cite{Bjornson4,Ramezani2022}
\begin{equation}
\begin{aligned}
\label{eq:array_gain_BD}
   &G_{\mathrm{array}}(0,0,z_r) \approx \left(\frac{8z_{\mathrm{eff}}}{MNd_{\mathrm{F}}} \right)^2 \cdot \\ &\left(C^2 \left(M \sqrt{\frac{d_{\mathrm{F}}}{8z_{\mathrm{eff}}}}\right)  + S^2 \left(M \sqrt{\frac{d_{\mathrm{F}}}{8z_{\mathrm{eff}}}}\right)\right)\cdot \\&\left(C^2 \left(N \sqrt{\frac{d_{\mathrm{F}}}{8z_{\mathrm{eff}}}}\right)  + S^2 \left(N \sqrt{\frac{d_{\mathrm{F}}}{8z_{\mathrm{eff}}}}\right)\right),
   \end{aligned}
\end{equation}
where $z_{\mathrm{eff}} = \frac{z_r F}{|z_r - F|}$ and  $C(x) = \int_{0}^{x} \cos(\pi t^2/2) dt$ and $S(x) = \int_{0}^{x} \sin(\pi t^2/2) dt$ are the Fresnel integrals. The array gain expression in \eqref{eq:array_gain_BD} takes its maximum value at $d_{\mathrm{F}}/8z_{\mathrm{eff}} = 0$ or equivalently $z_r = F$, which is the focal point.
To determine the $3\,$dB BD, we want to identify at what distance the array gain reduces to $0.5$. The array gain in \eqref{eq:array_gain_BD} is a function of $a =d_{\mathrm{F}}/(8z_{\mathrm{eff}})$ and we let $a_{3\,\mathrm{dB}}$ denote the value for which the array gain becomes $0.5$. This value must generally be obtained numerically.
We can then use the expression for $z_{\mathrm{eff}}$ to solve for $z_r$, which results in the two roots

\begin{equation}
\label{eq:3dBloss_depth}
    z_{r,3\,\mathrm{dB}} = \frac{d_{\mathrm{F}}F}{d_{\mathrm{F}} \pm 8a_{3\,\mathrm{dB}}F}.
\end{equation}
This equation reveals that there is a depth interval
\begin{equation} \label{eq:BD-intervals}
\frac{d_{\mathrm{F}}F}{d_{\mathrm{F}} + 8a_{3\,\mathrm{dB}}F} \leq z_r \leq \frac{d_{\mathrm{F}}F}{d_{\mathrm{F}} - 8a_{3\,\mathrm{dB}}F}
\end{equation}
for which the normalized array gain is between $0.5$ and $1$. The length of this interval is the $3\,$dB BD and it is obtained as 
\begin{align}
   \mathrm{BD}_{\mathrm{3\,dB}}  = \begin{cases}  \frac{16 a_{3\,\mathrm{dB}}d_{\mathrm{F}}F^2}{d_{\mathrm{F}}^2 - 64 a^2_{3\,\mathrm{dB}} F^2}, & F < \frac{d_{\mathrm{F}}}{8a_{3\,\mathrm{dB}}}, \\
    \infty,  & F \geq \frac{d_{\mathrm{F}}}{8a_{3\,\mathrm{dB}}},
    \end{cases} \label{eq:BD}
\end{align}
where the second case occurs when the upper limit in \eqref{eq:3dBloss_depth} becomes negative (i.e., non-existing).
According to \eqref{eq:BD}, the $3\;$dB BD is finite when the ELAA focuses on a point closer than $d_{\mathrm{F}}/8a_{3\,\mathrm{dB}}$ and extends to infinity when focusing on more distant points. For a square array with $M=N$, the BD expression in \eqref{eq:BD} is simplified as 
\begin{align}
   \mathrm{BD}_{\mathrm{3\,dB}}  = \begin{cases}  \frac{20d_{\mathrm{FA}}F^2}{d_{\mathrm{FA}}^2 - 100 F^2}, & F < \frac{d_{\mathrm{FA}}}{10}, \\
    \infty,  & F \geq \frac{d_{\mathrm{FA}}}{10}.
    \end{cases} \label{eq:BD_square}
\end{align}

\begin{figure}[t!]
       \centering
       \includegraphics[width=1\columnwidth]{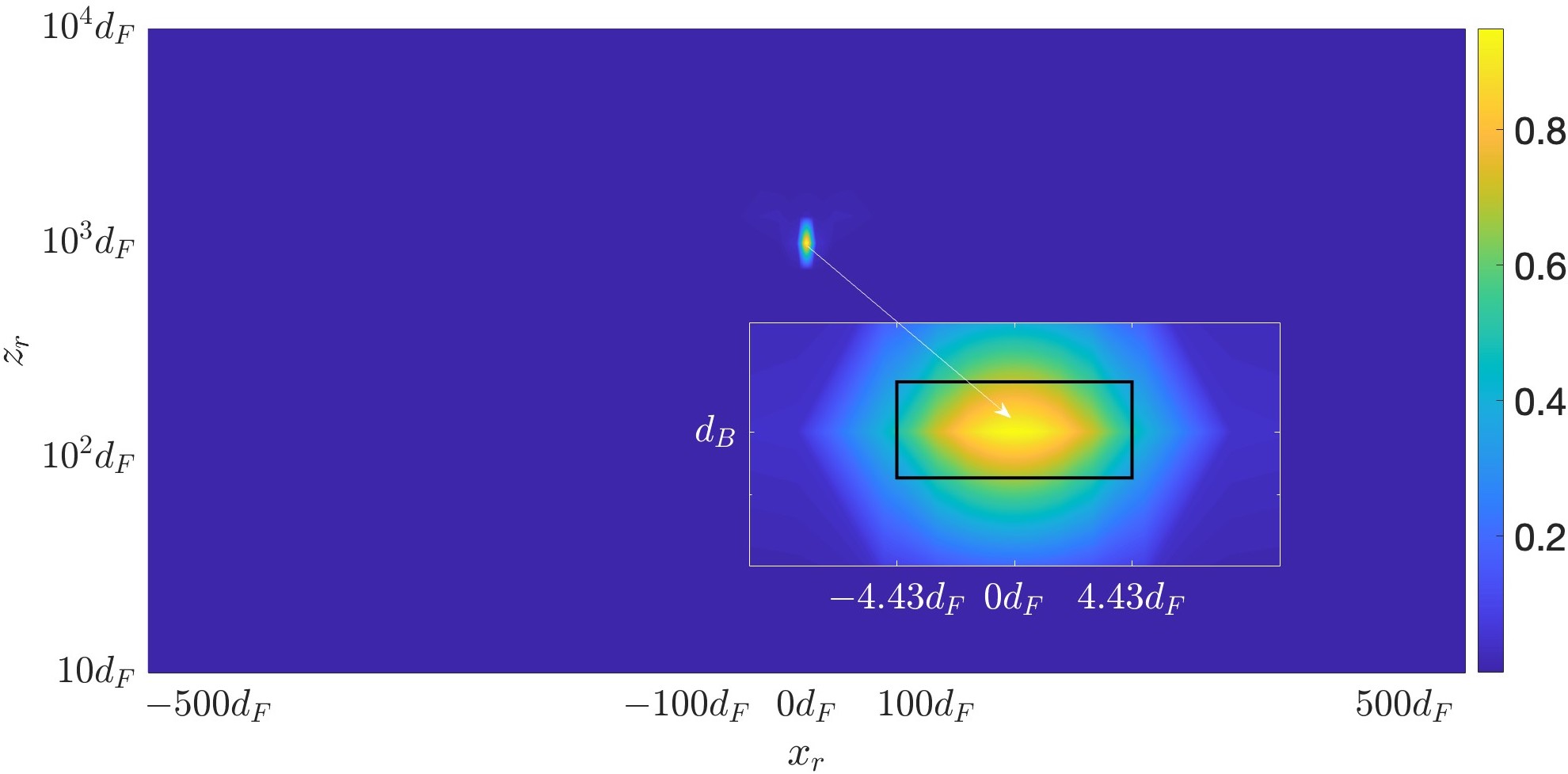}
       \caption{A heat map of the normalized gain when the ELAA focuses a signal at the point$(0,0,d_{\mathrm{B}}=1000\,d_{\mathrm{F}})$. Most of the beam energy is concentrated in the region characterized by the black rectangle.}
       \label{fig:heat_map}
   \end{figure}
   
To showcase how a beam focused at a near-field point behaves, Fig.~\ref{fig:heat_map} shows the heat map of the normalized gain when the ELAA (with the same configuration and number of antennas as in Fig.~\ref{fig:beam_width}) focuses at $(0,0,d_{\mathrm{B}})$. It can be observed that the beam energy radiated by the ELAA is concentrated in an area around the focal point with finite width and depth. 
Therefore, the beamforming in the radiative near-field only provides an array gain in a small region surrounding the focal point. This is an excellent new feature that enables 
 the massive spatial multiplexing paradigm that will be described next.

\section{Massive Spatial Multiplexing}
By utilizing the precise signal focusing ability that appears in the radiative near-field region, ELAAs can remarkably
improve the sum rate in future systems through spatial multiplexing of a massive number of data streams in both the depth and angular domains. This section will describe how this constitutes a paradigm shift in serving many users simultaneously and allows for sending multiple beams to one single user under LOS conditions. We will cover these two cases in separate sections.

\subsection{Massive Multiplexing in the Depth Domain}
We have demonstrated how beamforming in the radiative near-field generates beams with finite depth. This depth perception in the near-field allows the array to act like a lens that focuses the signal on a specific location instead of in a specific direction, as in the far-field. This enables a new multiplexing technique where multiple users that are positioned in the same angular direction with respect to the ELAA but at different distances can be simultaneously served since the channel vectors will be vastly different. Multiplexing in the depth domain is a game changer for serving massive crowds of users, which is hard for the BS to manage with traditional far-field beamforming.

\begin{figure}[t!]
       \centering
       \includegraphics[width=1\columnwidth]{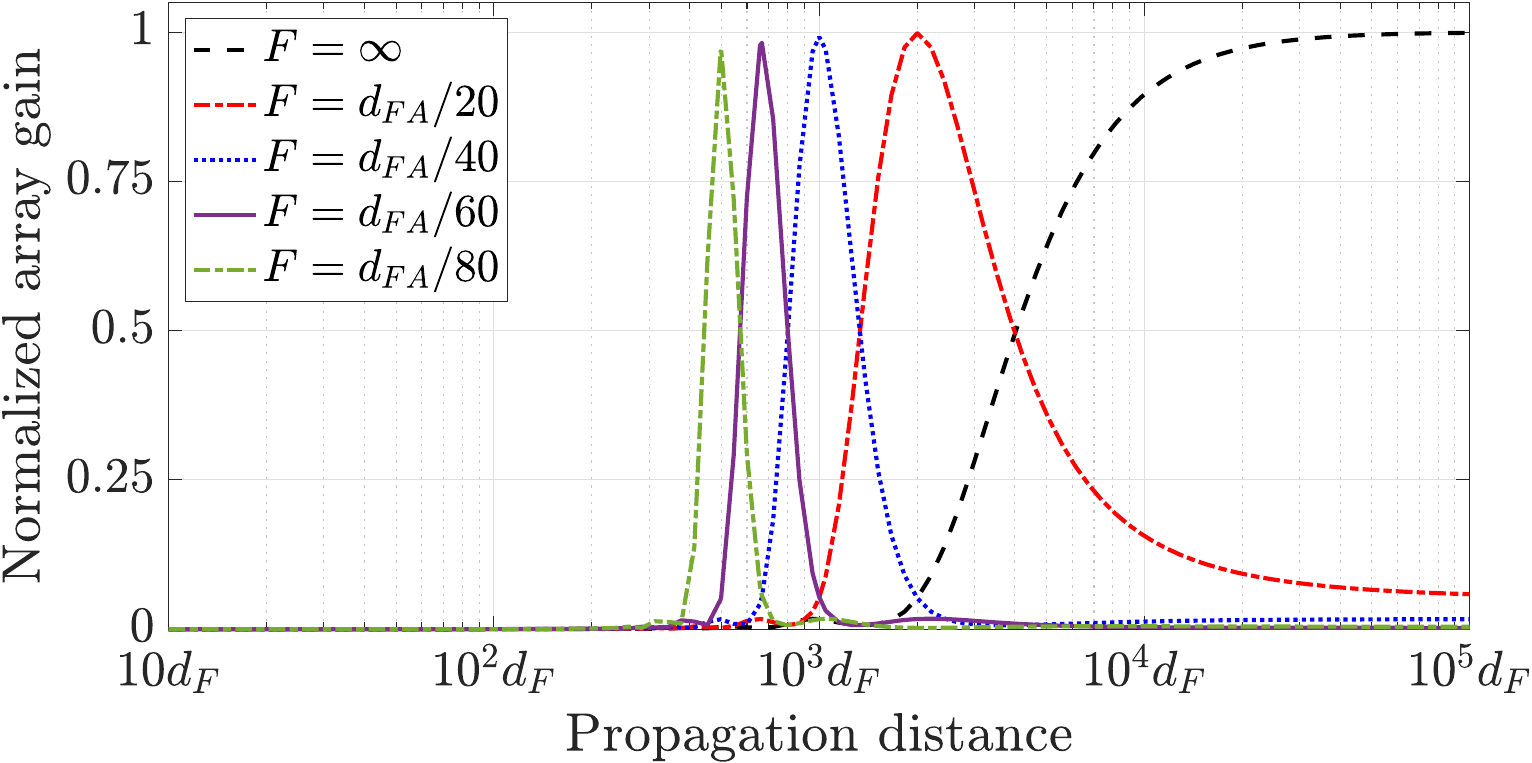}
       \caption{The normalized array gains are shown for five focal points in the same direction that have non-overlapping $3$\,dB BWs. This feature enables near-field multiplexing of these users in the depth domain with limited interference between the users.}
       \label{fig:nearfield_multiplexing}
       \vspace{-5mm}
   \end{figure}

According to \eqref{eq:3dBloss_depth}, we can have distinct BD intervals when focusing at distances closer than $d_{\mathrm{F}}/(8a_{3\,\mathrm{dB}})$. Specifically, for a square ELAA with $M=N$, we can utilize \eqref{eq:BD-intervals} to identify $F = \infty,\, F = d_{\mathrm{FA}}/20,\,F = d_{\mathrm{FA}}/40,F = \,d_{\mathrm{FA}}/60$, and $F =\,d_{\mathrm{FA}}/80$ as five focal points with non-overlapping $3$\,dB BD intervals. 
$F=d_{\mathrm{FA}}/32a_{3\,\mathrm{dB}}$, and so on. 
Fig.~\ref{fig:nearfield_multiplexing} shows the normalized array gain when an ELAA with $M=N=200$ and $D=\lambda/2$ focuses on the mentioned five points using matched filtering. It can be clearly seen that the $3\,$dB BD intervals are non-overlapping, which lets the ELAA transmit to the five users at the same time without causing much interference. Thus, the finite BD in the radiative near-field cater for the multiplexing of multiple users placed at the same angular direction but at different propagation distances.       

\begin{figure}[t!]
       \centering
       \includegraphics[width=1\columnwidth]{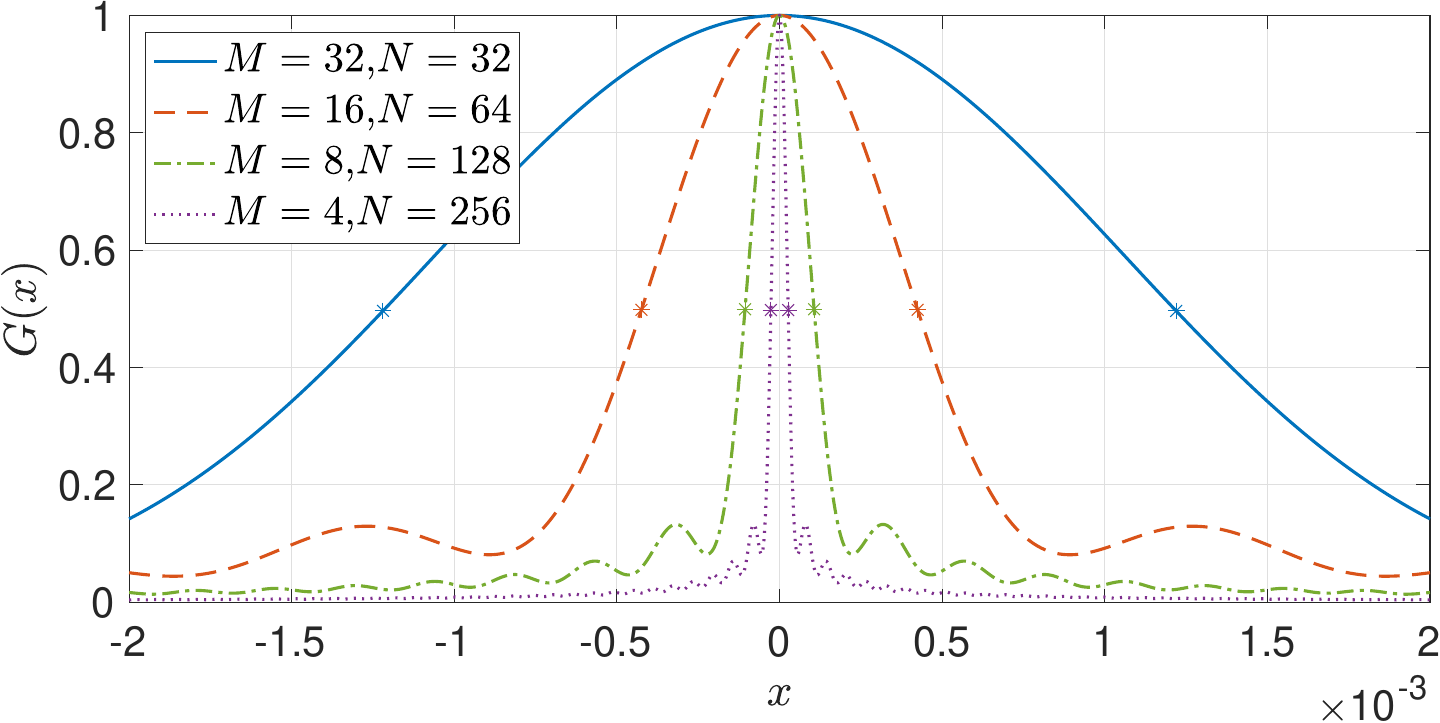}
       \caption{The function $G(x)$ in \eqref{eq:Gx-function} for different combinations of $M$ and $N$ that give $MN=1024$. The square configuration is widest and, therefore, provides the largest BD.}
       \label{fig:Gx}
       \vspace{-3mm}
   \end{figure}

The arrangement of the antennas in the array also plays a role in determining how many users can be spatially multiplexed. By taking a closer look at the BD expression in \eqref{eq:BD} for $F<\frac{d_{\mathrm{F}}}{8a_{3\,\mathrm{dB}}}$, we can notice that it is an increasing function of $a_{3\,\mathrm{dB}}$ which means that the BD expands if we increase $a_{3\,\mathrm{dB}}$.
For a fixed number of antennas, $a_{3\,\mathrm{dB}}$ is maximized when the array has a square shape with the same number of antennas in both dimensions and it becomes smaller as the difference between the number of horizontal and vertical antennas increases. To illustrate this, we consider four different arrays with the same total number of antennas $MN = 1024$ but different number of columns and rows. Fig.~\ref{fig:Gx} shows the array gain expression from \eqref{eq:array_gain_BD}, which we express as a function of $x = d_{\mathrm{F}}/8z_{\mathrm{eff}}$ as
\begin{align} \notag
&G(x) = \\ &\frac{\left(C^2(M\sqrt{x})+S^2(M\sqrt{x})\right)\left(C^2(N\sqrt{x})+S^2(N\sqrt{x})\right)
}{(MNx)^2}. \label{eq:Gx-function}
\end{align}
This function is shown for the four considered setups and, for the sake of illustration, it is plotted for both positive and negative values of $x$. The star markers in the figure indicate  $x = \pm\, a_{3\,\mathrm{dB}}$ for the corresponding graph. It can be observed that $|a_{3\,\mathrm{dB}}|$ reduces as the array shape changes from square to rectangle.

\begin{figure}
\centering
\subfloat[A square array with $M=N=200$.]
{\includegraphics[width=\columnwidth]{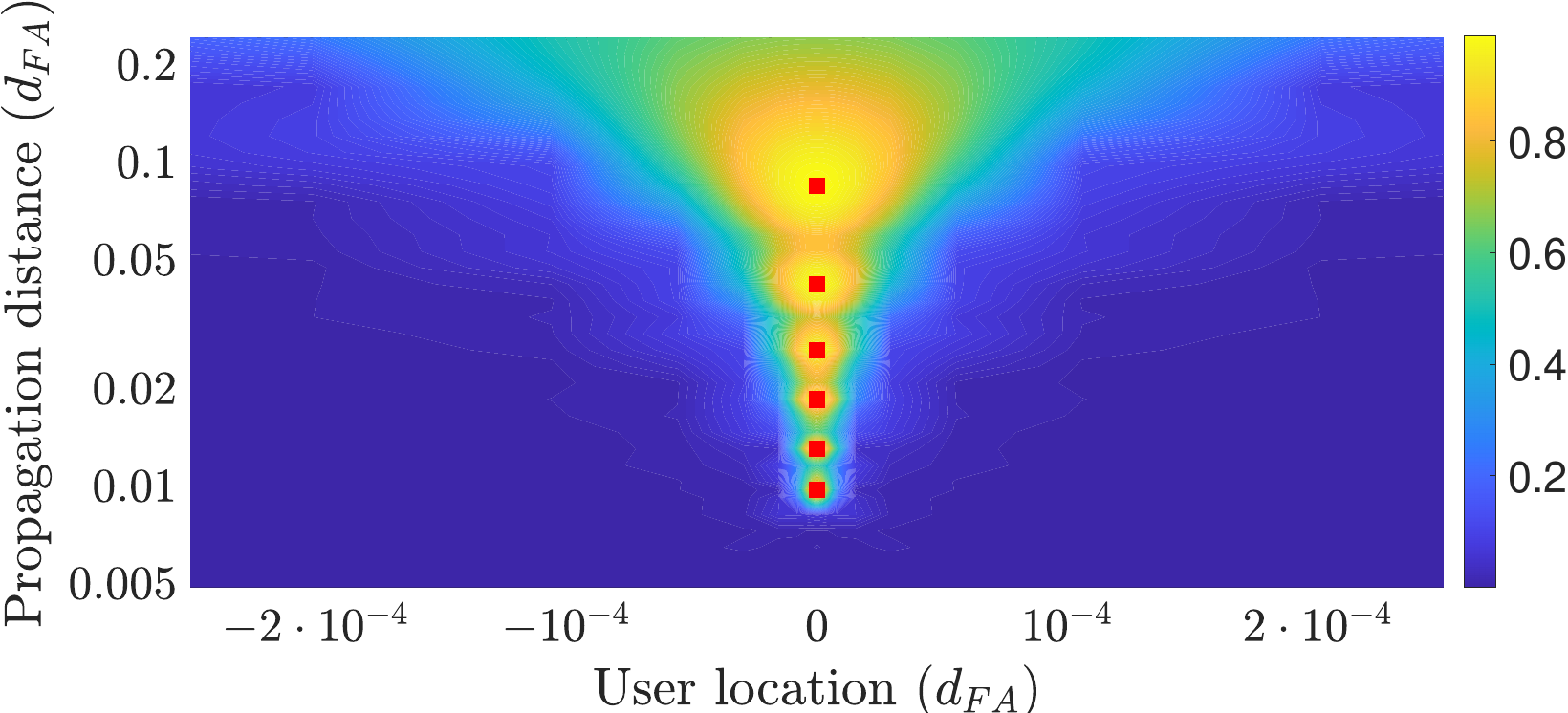}}\hfill
\centering
\subfloat[A rectangular array with $M=80$ and $N=500$.]
{\includegraphics[width=\columnwidth]{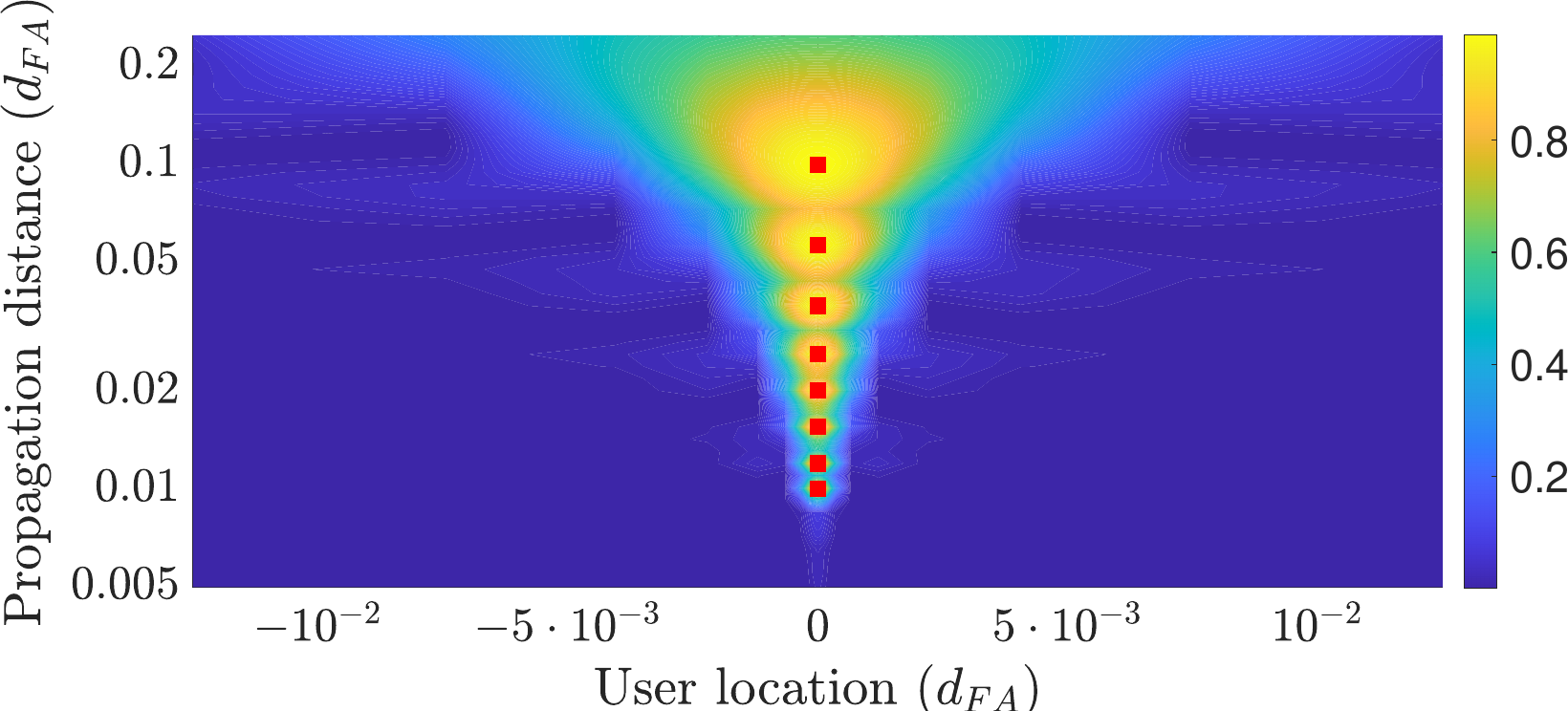}}
\caption{Multiplexing in the depth domain with an ELAA equipped with $4\cdot 10^4$ antennas, each with $D = \lambda/2$, using near-field beamforming. The ELAA is located at the origin $(0,0,0)$. The red dots indicate the focal points.}
\label{fig:multiplexing}
\vspace{-4mm}
\end{figure}

The above analysis indicates that a square array has a larger BD than a rectangular array having the same number of antennas. The smaller BD of a rectangular-shaped array allows for multiplexing more users in the depth domain. 
This can be observed in Fig.~\ref{fig:multiplexing} where the red dots show the served users. The transmitter has $MN = 4\cdot 10^4$ antennas. Fig. \ref{fig:multiplexing}(a) shows the scenario with a square array having $200$ antennas with $D=\lambda/2$ in each dimension,  while the array considered in Fig.~\ref{fig:multiplexing}(b) is rectangular with $M= 80,\,N=500$ antennas. For the rectangular case provides a smaller BD, which makes it possible  to multiplex more users in the depth domain than in the square case ($8$ vs. $6$ users in this example).   

Apart from these novel channel properties, the transmitter/receiver processing remains the same as in conventional MU-MIMO systems. For example, if $K$ single-antenna users are served and the channel vector to user $k$ is denoted as $\vect{h}_k \in \mathbb{C}^{MN}$, then the received downlink signal $y_k \in \mathbb{C}$ can be expressed as
\begin{equation} \label{eq:MU-MIMO-userk}
    y_k = \vect{h}_k^{\Htran} \sum_{i=1}^{K} \vect{w}_i x_i + n_k,
\end{equation}
where $x_i \in \mathbb{C}$ is the data signal intended for user $i$, $\vect{w}_i \in \mathbb{C}^{MN}$ is the corresponding precoding vector, and $n_k \in \mathbb{C}$ is the receiver noise at user $k$.
The combined received signals of all users can be expressed as
\begin{equation} \label{eq:MU-MIMO-userk}
    \underbrace{\begin{bmatrix}
        y_1 \\
        \vdots \\
        y_K
    \end{bmatrix}}_{=\vect{y}} = \underbrace{\begin{bmatrix}
        \vect{h}_1^{\Htran} \\ \vdots \\ \vect{h}_K^{\Htran}
    \end{bmatrix}}_{=\vect{H}^{\Htran}}  
    \underbrace{\begin{bmatrix}
        \vect{w}_1 & \ldots & \vect{w}_K
    \end{bmatrix}}_{=\vect{W}}
    \underbrace{\begin{bmatrix}
        x_1 \\
        \vdots \\
        x_K
    \end{bmatrix}}_{=\vect{x}} +     \underbrace{\begin{bmatrix}
        n_1 \\
        \vdots \\
        n_K
    \end{bmatrix}}_{=\vect{n}},
\end{equation}
which can be expressed as $\vect{y} = \vect{H}^{\Htran} \vect{W} \vect{x} + \vect{n}$.
The transmitter can then alleviate the interference by using the zero-forcing precoding matrix
\begin{equation}
    \vect{W} = \alpha \vect{H} (\vect{H}^{\Htran} \vect{H})^{-1},
\end{equation}
where $\alpha$ is a scaling factor that can be selected to satisfy a total power constraint. This precoding matrix is the pseudo-inverse of the channel matrix $\vect{H}^{\Htran}$, but it only exists if the channel matrix has rank $K$.
In a far-field scenario where the $K$ users are located in the same angular direction, all the rows in $\vect{H}^{\Htran}$ will be identical (apart from the scaling factor); thus, zero-forcing does not exist because the users cannot be resolved by the transmitter.
This is not the case in the radiative near-field, where the channel matrix is generated differently.

The improved spatial resolution that occurs in the radiative near-field is also useful in multipath propagation scenarios, where scattering clusters located at different distances can be resolved to make the channel vectors of different users more distinguishable than in the far-field.

\subsection{Massive Multiplexing in the Angular Domain}

\begin{figure}
\centering
\subfloat[Beamforming mode $1$ with $98.41 \%$ of the channel gain.]
{\includegraphics[width=\columnwidth]{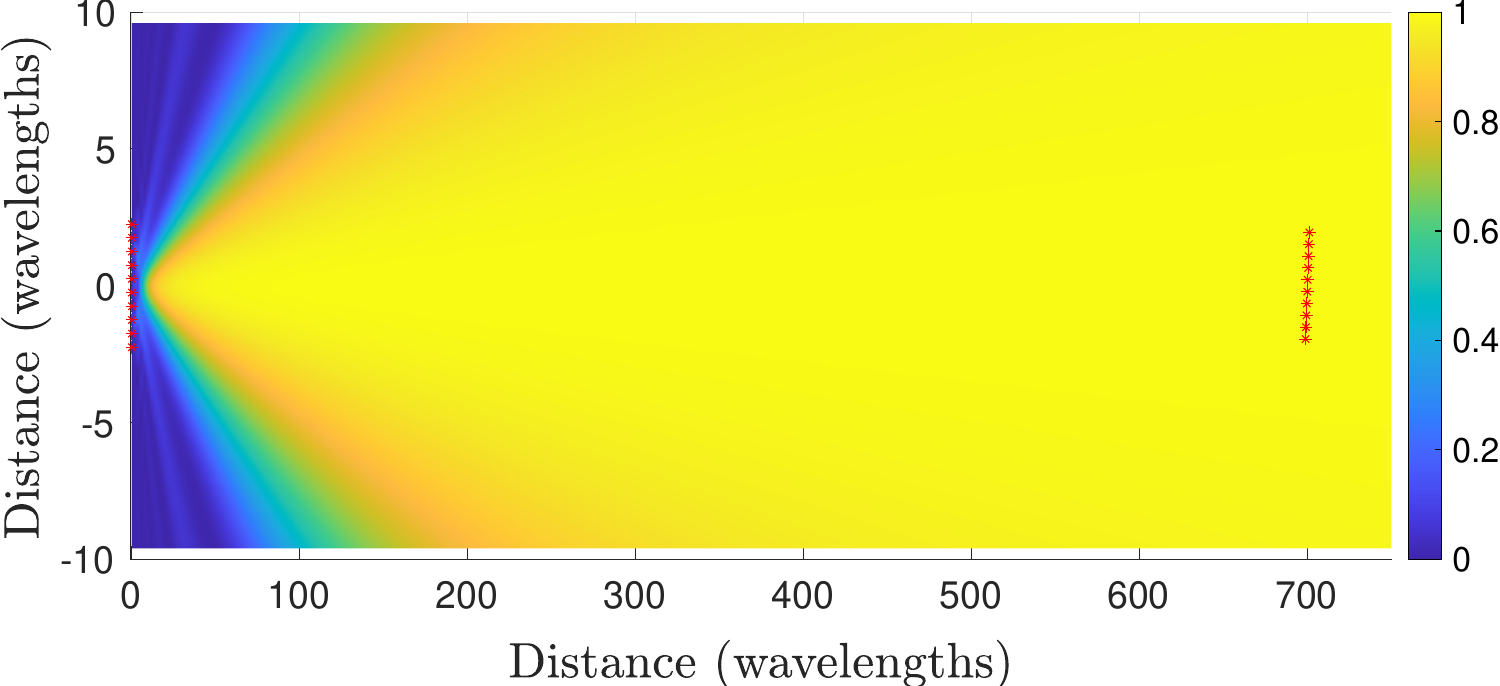}}\hfill
\centering
\subfloat[Beamforming mode $2$ with $1.58 \%$ of the channel gain.]
{\includegraphics[width=\columnwidth]{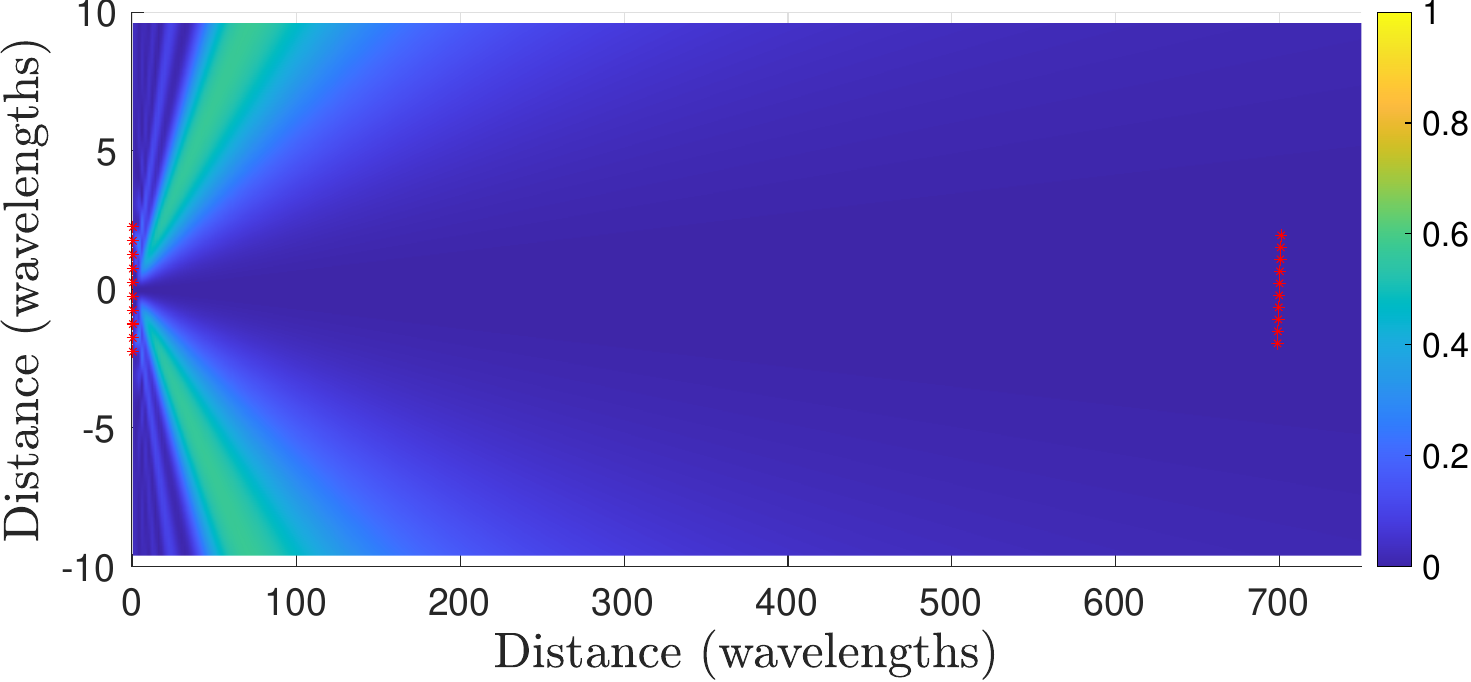}}
\caption{Far-field beamforming with $10$ transmit antennas and $10$ receive antennas in ULA configurations give a channel matrix for which $\vect{H}^{\Htran} \vect{H}$ only has one large eigenvalue. It corresponds to beamforming straight towards the receiver. The weak second eigenvalue beamforms in a completely different direction.}
\label{fig:rank_1}
\vspace{-4mm}
\end{figure}

We will now shift focus to SU-MIMO, for which the channel matrix $\vect{H}$ in free-space LOS scenarios is conventionally viewed to have rank $1$ \cite[Sec 7.2.3]{Tse2005a}. This is because the communication is assumed to occur in the far-field region, where the wavefronts are planar and the channel response is governed by a single dominant path. Fig.~\ref{fig:rank_1} exemplifies conventional far-field MIMO-LOS communication. In this case, $\vect{H}^{\Htran} \vect{H}$ has a single dominant eigenvalue containing the vast majority of the channel gain.
The corresponding dominant beamforming mode, shown in Fig.~\ref{fig:rank_1}(a), corresponds to beamforming straight towards the receiver and placing it in the middle of a wide beam. However, there are always additional weak beamforming modes, such as the one shown in Fig.~\ref{fig:rank_1}(b), that contains $1.5\%$ of the channel gain.

\begin{figure}[t!]
    \centering
           \begin{overpic}[width=\columnwidth,tics=10]{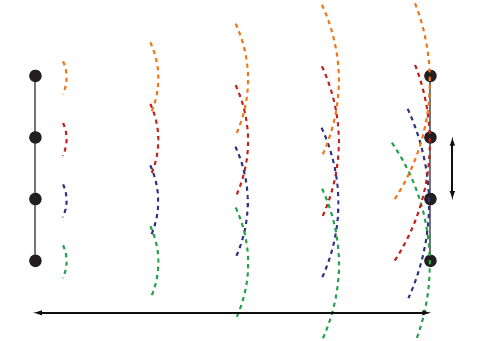}
  \put(2.4,54){\small $1$}
\put(2,15.5){\small $K$}
  \put(91,54){\small  $1$}
  \put(91,15.5){\small $K$}
   \put(94.5,33.5){\small $\Delta$}
   \put(48,1.5){\small $d$}
\end{overpic}  \vspace{-4mm}
    \caption{An example of SU-MIMO communication in the radiative near-field where the receiver can distinguish the transmitted signals based on their spherical curvatures.}
    \label{fig_multi_spherical}
\end{figure}
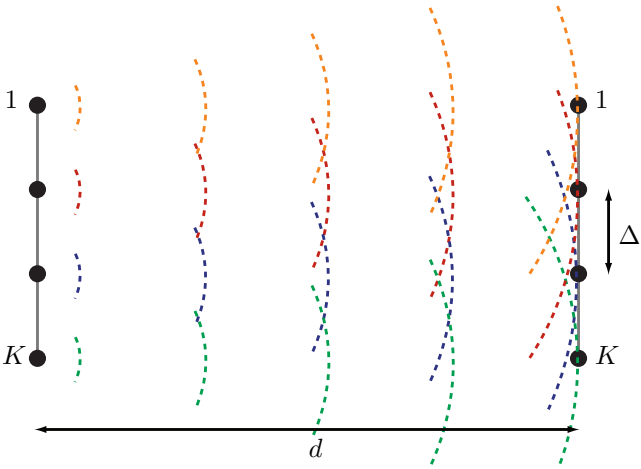

In the radiative near-field, we can make use of the spherical wavefront characteristics to achieve ideal full-rank MIMO channels with equal eigenvalues also in LOS scenarios. Fig.~\ref{fig_multi_spherical} sketches the signal transmission between two ULAs with $K$ antennas. We can see that the receiver can distinguish between the signals transmitted from the different antennas based on their spherical curvatures. This enables the receiver to distinguish multiple signals that are transmitted in LOS, particularly if the antenna spacing $\Delta$ is properly selected.

We will now delve deeper into the free-space LOS communication between two ULAs. The arrays are in the broadside direction and separated by a distance $d$. They are identically arranged with $K$ antennas.
The antenna spacing is $\Delta$, as illustrated in Fig.~\ref{fig_multi_spherical} \cite{Lu2023}.
The distance between the $k$th transmit and $m$th receive antenna, depending on the antenna spacing, is 
\begin{equation}
d_{m,k}=\sqrt{d^2+\big( m-k\big)^2 {\Delta}^2},  
 \end{equation}
 where $m,k \in \{1,\dots, K\}$. Therefore, we can model the MIMO channel for a LOS scenario as \cite{Torkildson}
\begin{equation}\label{eq:Hu}
\mathbf{H}=    
\begin{bmatrix}
\sqrt{\beta_{1,1}} e^{-\imagunit\pi\frac{d_{1,1}-d}{\lambda}} & \cdots & \sqrt{\beta_{1,K}}e^{-\imagunit\pi\frac{d_{1,K}-d}{\lambda}}\\
\vdots & \ddots & \vdots \\
\sqrt{\beta_{K,1}}e^{-\imagunit\pi\frac{d_{K,1}-d}{\lambda}} & \cdots & \sqrt{\beta_{K,K}}e^{-\imagunit\pi\frac{d_{K,K}-d}{\lambda}}\\
\end{bmatrix},
\end{equation}
where the phase-shifts are determined by the distance $d_{m,k}$ between different transmit and receive antenna pairs  and the reference distance $d$. The channel gain between the transmit and receive antennas is 
\begin{equation}
\beta_{m,k} =  G_t G_r \left( \frac{ \lambda}{4 \pi d_{m,k}} \right)^2,
\end{equation}
where $G_t$ and $G_r$ are the transmit and receive antenna gains, respectively. In the case of isotropic antennas, they are both equal to one.
The diagonal in each array has a length of $  D =\left(K -1 \right) \Delta  $.
In typical propagation scenarios, the channel gain is nearly the same between all antenna locations beyond $d_{\mathrm{B}}$:
\begin{equation}
\beta_{m,k} \approx \beta = \left(\frac{\lambda}{4 \pi d}\right)^2.
\end{equation}
Furthermore, we can approximate the distance between the transmit and receive antennas using the first-order Taylor approximation as  $d_{m,k} \approx d + \frac{\delta_{m,k}} {2d}$, where $\delta_{m,k} = \left(m-k \right)^2 {\Delta}^2  $. Based on this Fresnel approximation, the MIMO channel matrix in \eqref{eq:Hu} can be rewritten as 
\begin{align}\label{eq:Hu_approx}
\mathbf{\tilde{H}}&\approx \sqrt{\beta}
\begin{bmatrix}
e^{-\imagunit\pi\frac{\delta_{1,1}}{d\lambda}} & \cdots & e^{-\imagunit\pi\frac{\delta_{1,K}}{d\lambda}}\\
\vdots & \ddots & \vdots \\
e^{-\imagunit\pi\frac{\delta_{K,1}}{d\lambda}} & \cdots & e^{-\imagunit\pi\frac{\delta_{K,K}}{d\lambda}}\\
\end{bmatrix}.
\end{align}
It follows that $\|\mathbf{\tilde{H}}\|_F^2=\beta K^2 $ is the sum of eigenvalues of $\mathbf{\tilde{H}}^H \mathbf{\tilde{H}}$. As pointed out at the beginning of this article, it is preferable if all the eigenvalues (i.e., the squared singular values of $\mathbf{\tilde{H}}$) are equal. We can achieve this by tuning the antenna spacing $\Delta$ \cite{SongX2022}.
An arbitrary off-diagonal entry  $(k,\ell)$th element of $\mathbf{\tilde{H}}^H \mathbf{\tilde{H}}$ where $\ell \in \{1,\dots,K \}$ and $\ell \neq k$, have the magnitude
\begin{equation}\label{eq_off_diag}
    \beta \left| \sum_{m=1}^{K} e^ {\frac{\imagunit 2 \pi}{d \lambda} m \left(k - l\right) {\Delta}^2}   \right| =  \beta \left| \frac{1- e^{\imagunit \pi \frac{ 2 K \left( l - k \right) {\Delta}^2 }{\lambda d} }  }   {1- {e^{j \pi \frac{2\left( l - k \right) {\Delta }^2 }{\lambda d}} }}  \right|.
\end{equation}
 The last equation is obtained by using the classical sum of geometric series \cite{DoHeedong2020ColM}.
If we have $\frac{K {\Delta}^2}{\lambda d} =1$,
then the magnitude of all off-diagonal entries is $0$. In this case, all the $K$ eigenvalues are equal to $\beta K$ since  $\mathbf{\tilde{H}}^H \mathbf{\tilde{H}} = \beta K \mathbf{I}$ and the capacity is maximized. Therefore, we obtain the capacity-maximizing antenna spacing as \cite{Torkildson,renzo2023}
\begin{equation}
    {\Delta} =  \sqrt{\frac{\lambda d}{K}}.
\end{equation}
The capacity with the optimal spacing results in  the upper bound in \eqref{eq:upper_limit}, i.e.,  
\begin{equation}\label{capacityusa}
    C= B K \log_2 \left(1+\frac{P  \beta  }{B N_0} \right).
\end{equation} 
Notice that the beamforming gain inside of the SNR term is canceled since we distribute the power equally across users.
The capacity can be enhanced by increasing the number of antennas $K$, which simultaneously increases the multiplexing gain and beamforming gain.

The area of the array is $A_\textrm{array} = (  \sqrt{\frac{\lambda d}{K}} (K-1)+ \tilde{W} )^2$, where $\tilde{W}\geq 0$ is the width of an antenna. For a fixed array area $A_\textrm{array}$ and given distance $d$ between the transmit and receive arrays, the multiplexing gain $K$ is a function of the wavelength $\lambda$ through $\beta$, which can be tightly approximated by $A_\textrm{array}^2/(\lambda d)^2$ when $\lambda$ is small. More specifically, the multiplexing gain is inversely proportional to the wavelength, implying a preference for using higher carrier frequencies to achieve a higher multiplexing gain.

\begin{figure}
    \centering
    \includegraphics[width=\columnwidth]{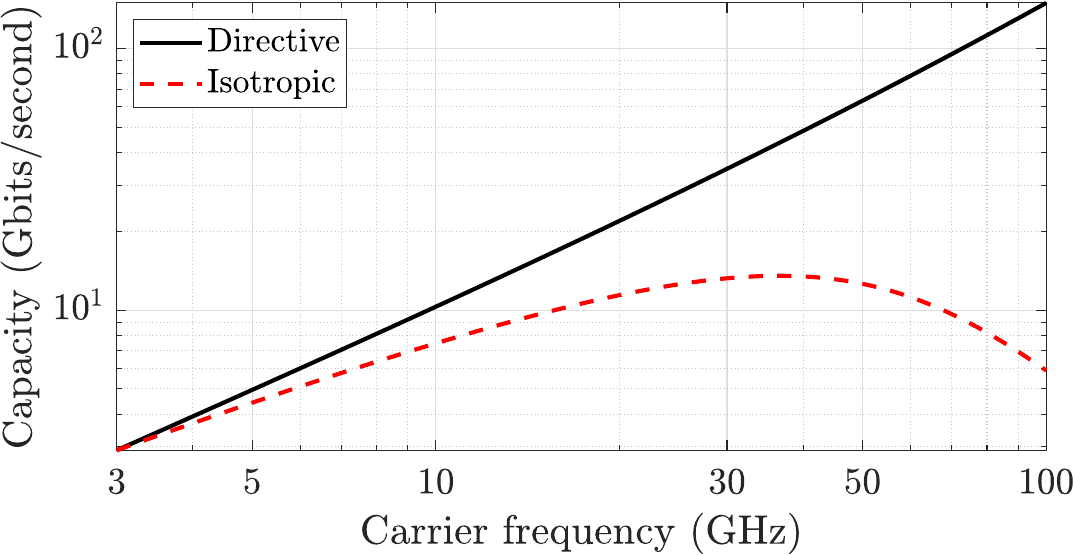}
    \caption{The capacity with respect to the wavelength. 
    In the case of isotropic antennas, $G_t = G_r = 1$; In the case of directive transmitter antennas, $G_t = G_r = 1/\lambda$.   }
    \label{fig_C_vs_freq}
    \vspace{-4mm}
\end{figure}

We evaluate the capacity expression in \eqref{capacityusa} with respect to the
carrier frequency in Fig.~\ref{fig_C_vs_freq}. We select a fixed array area  $A_\textrm{array}=0.01 \textrm{ m}^2 $,  distance $d=10$ m, $\tilde{W}=\lambda/2$,  $P/N_0=189.03$ dB, and $B=0.03f_c$, where $f$ denotes the carrier frequency. First, we consider isotropic antennas at both the transmitter and receiver, i.e., $G_t=G_r=1$. In that case, $\beta$ is proportional to $\lambda^2$. The capacity with respect to the carrier frequency is illustrated by the red curve in Fig.~\ref{fig_C_vs_freq}. We observe that the capacity increases for carrier frequencies less than $40$ GHz. This can be attributed to the multiplexing gain outweighing the SNR reduction. However, as we keep on increasing the carrier frequency, the capacity reaches a plateau and then decreases (as in Fig.~\ref{fig:convergence_upper_bound}). In contrast, if we consider directive antennas (i.e., by setting $G_t = G_r = 1/\lambda$), we can keep the SNR constant as the frequency changes. The black curve in Fig.~\ref{fig_C_vs_freq} depicts the corresponding capacity. The capacity improves significantly compared to the case of isotropic antennas. For instance, with a carrier frequency of $70$ GHz, the capacity improves approximately one order of magnitude.
The analysis above reveals several intriguing features that are at the heart of massive spatial multiplexing:
\begin{enumerate}
    \item Increasing the carrier frequency for the purpose of solely utilizing greater bandwidth does not guarantee capacity enhancement, as the SNR diminishes with short wavelengths. This finding is consistent with the observation in Fig.~\ref{fig:convergence_upper_bound}, where the capacity (measured in bit/s) reaches its upper limit as the bandwidth increases. 
    \item The beamforming gain does not enhance the capacity as it gets nullified by the equal distribution of power. Therefore, equation \eqref{capacityusa} disregards the presence of the beamforming gain term.
    \item 
    The multiplexing gain consistently enhances the capacity, which serves as a driving force behind our proposition of a new paradigm. This paradigm revolves around seeking the multiplexing gain achieved through the utilization of additional antennas.  Thanks to the progressive trend towards higher frequency in future wireless communications, enabling the incorporation  of more antennas within the same array's area.
\end{enumerate}

\section{Research Challenges and Future Directions}

In this section, we explore future research challenges and directions related to  massive spatial multiplexing. We categorize them into four categories as follows.

\subsection{Signal Processing Tailored to the Near-Field}

    The spherical wavefronts in near-field communications introduce the depth domain as a novel degree of freedom that enables massive spatial multiplexing. The channel matrix $\vect{H}$ must be estimated to achieve these gains, which is associated with a high complexity if traditional non-parametric methods (e.g., least squares) are used. A potential solution is to tailor the signal processing algorithms to the near-field scenario by utilizing channel parametrizations that capture the depth. There are major challenges related to this as well; for example, the channel estimation and beamforming codebook design must account for the additional depth grid \cite{Lu2023}, otherwise the ability to distinguish signals arriving from similar angular directions is lost. 
    There is also a risk that limited phase-synchronization, manufacturing errors, and the presence of non-linear phase characteristics will further limit the signal processing performance. Therefore, it is required to study the redesign of the signal processing algorithms in detail.
    
    Deep learning (DL) techniques might pave the way to overcome these challenges. DL techniques can be utilized to design a near-field codebook that accounts for non-linear phase shifts by leveraging the inherent non-linearity of neural networks \cite{LiuLowOverhead}. To overcome the channel estimation challenge, a deep learning-based solution is reported in \cite{Zhang2023}. The paper formulates the estimation as a compressed sensing problem and then uses a model-driven learning approach to solve it. In \cite{Zhang2023DeepLearning}, a DL approach is proposed to perform frequency-aware beamforming that prevents the beamforming gain degradation due to the wideband effect. Despite these recent advances,  unlike the DL applications in far-field communications that are well studied, further investigations are necessary to fully understand and optimize the application of DL techniques in the near-field communications.
When effective learning-based solutions have been developed, it is also desirable to extract the key properties for improved modeling.

\subsection{Practical Hardware Models}

For theoretical studies to be applicable to practical scenarios, accurate and realistic hardware models must be used, which take into account the intrinsic properties of ELAAs. For instance, the mutual coupling effects that have traditionally been neglected by the communication community (e.g., motivated by placing antennas half a wavelength apart) must be included in the system models to obtain physically accurate results. 
Coupling-aware communication models are needed to accurately characterize the coupling between adjacent antennas, particularly, in arrays with densely-packed antenna elements. Though mutual coupling has been known to degrade the capacity in MIMO systems, it has been recently shown that the effective utilization of this property can also lead to high array gains in particular directions \cite{Williams2020}.

\subsection{Approaching the Spatial Degrees of Freedom}

The Nyquist-Shannon sampling theorem determines how many orthogonal channels an array of a given size can distinguish between. 
This number is called the \emph{spatial degrees of freedom} \cite{Poon}. For a planar array with a given area $A_\textrm{array}$, the degrees of freedom is \cite{Hu}
\begin{equation} \label{eq:spatialDOFs}
\textrm{Degrees of freedom} = \pi \frac{A_\textrm{array}}{\lambda^2}.
\end{equation}
The principle is that $\pi$ channels can be distinguished per area component of size $\lambda^2$. The number in \eqref{eq:spatialDOFs} is much larger than the number of signals that were spatially multiplexed in the SU-MIMO case, because the receiver only ``fills'' a small fraction of the transmitter's field of view.
In other words, we have only scratched on the surfaces of massive spatial multiplexing; we can combine the SU-MIMO and MU-MIMO cases to achieve even more effective multiplexing.

One open research question is which array geometry and antenna distribution within the aperture can make the most effective use of the spatial degrees of freedom. In addition to uniform planar arrays, as considered in this article, another classical configuration is uniform circular arrays (UCAs) \cite{Edfors2012a}. A high-rank LOS channel matrix can also be achieved when two UCAs are placed in the radiative near-field of each other. An interesting side-effect is that the ``beams'' take the shape of orbital angular momentum (OAM) modes, which emphasizes how different array geometries lead to different channel parametrizations and MIMO matrices.
The choice of antenna spacing is another open question: are there are any tangible gains from using shorter spacings than $\lambda/2$?

\subsection{Massive Spatial Multiplexing with RIS}

\begin{figure}[t!]
	\centering
	\begin{overpic}[width=1\columnwidth,tics=10]{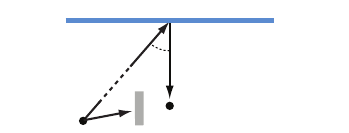}
	 \put (1,2.5) {Transmitter}
	 \put (51,6) {Receiver}
	 \put (43.5,20.5) {$\theta$}
	 \put (22,34) {Extremely large aperture RIS}
\end{overpic}  \vspace{1mm}
	\caption{In an RIS-aided communication scenario where both the transmitter and receiver are in the near-field region of the surface, a high-rank virtual LOS path can be created between them.}
	\label{fig:RIS_setup}
 \vspace{-4mm}
\end{figure}
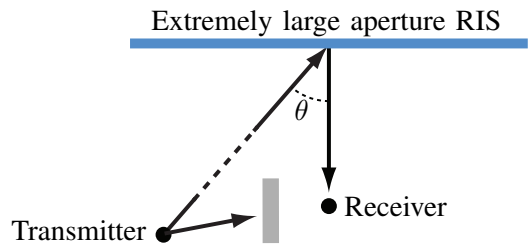

The near-field characteristics also influence communication systems that involve reconfigurable intelligent surface (RIS). These surfaces were conventionally viewed as reflectors that can create a virtual rank-one LOS path between a transmitter-receiver pair\cite{Bjornson1}. Such a scenario is illustrated in Fig.~\ref{fig:RIS_setup}, where the direct path between the transmitter and receiver is blocked but signals can be reflected off the RIS. If both the transmitter and receiver are in the radiative near-field, the RIS can create a high-rank channel between the transmitter and receiver \cite{Do2023a}.
The theory described earlier in this paper can be utilized to compute the $3\,$dB BW and BD of the beam produced by the RIS  \cite{Bjornson4}. Consequently, for an \emph{extremely large aperture RIS} with the physical characteristics as an ELAA (i.e., the same number of elements and element spacing in the RIS as the number of antennas and antenna spacing in the ELAA), the $3$\,dB BW and BD are equal to those of the ELAA. Hence, all the previously mentioned research challenges can also be explored in the context of RIS-aided communications. Moreover, phase-dependent amplitude variations of the elements should be considered when designing the RIS configuration.

\section{Conclusions}

The relentless pursuit for elevated data rates in forthcoming wireless technologies remains insatiable. However, simply increasing the bandwidth is inadequate, as data rates eventually reach a plateau. The focus should shift towards transmitting multiple parallel data streams as spatial layers in both SU-MIMO and MU-MIMO scenarios. In the former case, large arrays are employed on both the transmitter and receiver sides enabling angular-domain multiplexing of data streams to a single user, whereas in the latter scenario, an ELAA is utilized at the BS allowing for depth-domain multiplexing of multiple users. 
In particular, in the depth domain, simultaneous multiple user transmissions in the radiative near-field region are enabled using the signal focusing property. In the angular domain, spherical wavefront properties are leveraged to achieve an ideal full-rank MIMO channel with equal eigenvalues in LOS scenarios. Our numerical examples have showcased a consistent improvement in data rate by utilizing the benefits of multiplexing gain. This allows us to assert that we are now entering a new era marked by the advent of massive spatial multiplexing. Finally, we have posted some open issues and future directions such as the redesign of the signal processing aspect possibly using deep learning, the need for practical hardware considerations, exploration of different array shapes to approach the spatial degrees of freedom limit, and investigation of massive spatial multiplexing in RIS-aided communication.

\bibliographystyle{IEEEtran}
\bibliography{refs}

\end{document}